\documentclass[twocolumn,prb,showpacs,superscriptaddress]{revtex4}
\usepackage{epsfig}
\usepackage{amssymb}
\usepackage{dcolumn}
\usepackage{bm}
\begin{document}
%
\title{Hysteresis in a system driven by either generalized force or displacement
  variables}
\author{Erell Bonnot}
\affiliation{ Departament d'Estructura i Constituents de la Mat\`eria,
  Universitat de Barcelona \\ Diagonal 647, Facultat de F\'{\i}sica,
  08028 Barcelona, Catalonia}
\author{Ricardo Romero}
\affiliation{ Departament d'Estructura i Constituents de la
  Mat\`eria,  Universitat de Barcelona \\
  Diagonal 647, Facultat de F\'{\i}sica, 08028 Barcelona, Catalonia}

\affiliation{IFIMAT. CICPBA and Universidad Nacional del Centro.
Pinto 399, 7000 Tandil, Argentina}

\author{Xavier Illa}
\affiliation{ Departament d'Estructura i Constituents de la
  Mat\`eria,  Universitat de Barcelona \\
  Diagonal 647, Facultat de F\'{\i}sica, 08028 Barcelona, Catalonia}
\author{Llu\'{\i}s Ma\~nosa}
\affiliation{ Departament d'Estructura i Constituents de la Mat\`eria,
  Universitat de Barcelona \\ Diagonal 647, Facultat de F\'{\i}sica,
  08028 Barcelona, Catalonia}
\author{Antoni Planes}
\affiliation{ Departament d'Estructura i Constituents de la Mat\`eria,
  Universitat de Barcelona \\ Diagonal 647, Facultat de F\'{\i}sica,
  08028 Barcelona, Catalonia}
\author{Eduard Vives}
\email{eduard@ecm.ub.es}
\affiliation{ Departament d'Estructura i Constituents de la Mat\`eria,
  Universitat de Barcelona \\ Diagonal 647, Facultat de F\'{\i}sica,
  08028 Barcelona, Catalonia}
\date{\today}
\begin{abstract}
We report on experiments aimed at comparing the hysteretic
response of a Cu-Zn-Al single crystal undergoing a
martensitic transition under {\it strain-driven} and {\it
stress-driven} conditions. {\it Strain-driven} experiments were
performed using a conventional tensile machine while a special
device was designed to perform {\it stress-driven} experiments.
Significant differences in the hysteresis loops were found. The
{\it strain-driven} curves show re-entrant behaviour (yield point)
which is not observed in the {\it stress-driven} case. The
dissipated energy in the {\it stress-driven} curves is larger than
in the {\it strain-driven} ones. Results from recently proposed
models qualitatively agree with experiments.
\end{abstract}

\maketitle

\section{Introduction}

Hysteresis is ubiquitous in many areas of physics.  It is a signature of non-equilibrium effects and has attracted the attention of researchers for
a long time and is still one of the most challenging subjects of research \cite{Bertotti06}. Typically, the phenomenon of hysteresis results in
closed loops when plotting the external driving force (or field) {\it vs.} the generalized displacement (or vice versa). These two variables
(external force and generalized displacement) are conjugate variables in the first principle of thermodynamics and depending upon the specific
system, they can have different tensorial character. Nevertheless, in all cases their product renders the work performed on the studied system and
the area enclosed within a closed loop gives the energy loss.

Among hysteretic systems, particular attention has been devoted to athermal cases \cite{Sethna2005}. In these cases, hysteresis is not due to
competition between the fast driving rate and slow thermal relaxation, but rather it originates from the existence of very high energy barriers that
can only be overcome when the system reaches local marginal stability limits \cite{Cao1990,Perez2001}. The continuous change in the driving force
results in discontinuous changes in the conjugate displacement, giving rise to avalanche dynamics. Such dynamics has been observed in a wide variety
of materials such as ferromagnetic \cite{Puppin2000}, ferroelectric \cite{Colla02} and martensitic \cite{Vives1994} materials; field-driven vortex
motion in type-II superconductors \cite{Altshuler2004}, and pressure-driven condensation of $^4$He on mesophorous solids \cite{Lilly1993}, among
others. Since most of these systems show common hysteretic behaviour they are catalogued under the name of ferroic materials \cite{Wadhawan2000}.
They display first order phase transitions with an order parameter (exhibiting discontinuous behaviour), which corresponds to the generalized
displacement. A common feature in these systems is the existence of long range effects arising from compatibility conditions.

From an experimental point of view, it is usually easier to control the force while the conjugate displacement is measured. For instance, in
magnetic systems, the magnetic field is readily controlled and magnetic flux is measured. Although more difficult, it has also been possible to
control the magnetic flux \cite{Grosse1977} - which is suitable for studying materials displaying a rapid increase of magnetization. Moreover, the
control of the amount of gas (generalized displacement) instead of the chemical potential has been shown to be the most convenient procedure in a
number of gas adsorption experiments \cite{Wong1990}.

For a macroscopic system in equilibrium, thermodynamic trajectories do not depend on which is the control variable (generalized force or
displacement) because the two cases are related by a Legendre transformation \cite{Huang1987}. However, there is no reason to expect that out of
equilibrium hysteresis loops obtained when controlling the force will be similar to those obtained when the control parameter is the generalized
displacement. Such a situation has been very recently theoretically studied \cite{Illa2006a,Illa2006b} by making use of the Random Field Ising Model
(RFIM) \cite{Sethna1993}, which is a prototype model for athermal hysteresis. Results predict significant differences in the hysteresis loops
obtained in the force-driven case (external field $H$) to those obtained by driving the generalized displacement (magnetization $M$). It is the aim
of the present paper to address this issue from an experimental point of view.

Martensitic transitions (MT) offer a unique scenario to undertake such a task. A MT is a displacive first-order transition which involves a change
in symmetry \cite{Nishiyama1978}. In these ferroic systems, the transition can be induced by cooling but also by application of a mechanical stress.
In this latter case, the driving force is the applied load, while the conjugate displacement is the elongation (or strain). Typical experiments are
carried out using commercial tensile testing machines in which the control variable is the elongation\cite{feedback}. We have developed an
experimental device which enables fine control of the applied force while the strain is monitored. In this paper we present experiments performed on
the same specimen under both force and displacement control conditions. Results will enable a meaningful comparison of the hysteresis loops obtained
in the two cases. In addition, comparison with the predictions of recent theoretical approaches \cite{Illa2006a,Illa2006b} will also be presented.

\section{Experimental}

A Cu-Zn-Al single crystal was grown by the Bridgman technique. The nominal composition was chosen so that the (athermal) MT on cooling without
stress takes place from a cubic (L2$_1$) to a monoclinic ($18R$) structure slightly below room temperature (T$_M$ = 234K). The actual composition,
obtained from electron dispersion analysis is Cu$_{68.13}$Zn$_{15.74}$Al$_{16.13}$. A sample was mechanically machined from the ingot with
cylindrical heads. The body of the sample has flat faces  35 mm long, 1.4 mm thick and 3.95 mm wide. The axis of the sample is close to the [001]
crystallographic direction of the cubic phase. The sample was mechanically polished and then annealed for 20 min at 1073K, cooled in air down to
room temperature and aged for 2 h in boiling water. This heat treatment ensures that the sample is in the ordered state, free from internal stresses
and that the vacancy concentration is minimum at room temperature.

\begin{figure}[htb]
\begin{center}
  \epsfig{file=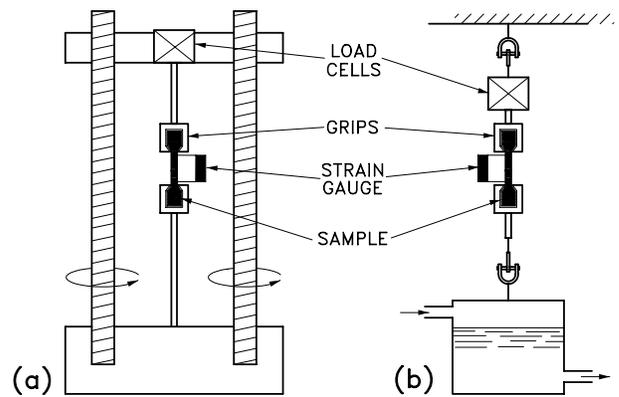,width=8.0cm,clip=}
\end{center}
\caption{\label{FIG1} Schematic representation of the strain-driven (a)
and stress-driven (b) experimental devices}
\end{figure}

Special grips, which adapt to the heads of the specimen, can be used in both experimental devices. In the first device [Fig. \ref{FIG1}(a)], a
INSTRON 4302 tensile machine, (from now on {\it strain-driven}), the control parameter is the elongation while the load is continuously monitored. A
second device (from now on {\it stress-driven}) was especially designed which enables control of the load applied to the sample while the elongation
is continuously monitored [see Fig. \ref{FIG1}(b)]. The device was adapted from that previously used by Carrillo et al. \cite{Carrillo97}.  The
upper grip is attached to a load cell hanging from the ceiling. Upper and lower ball-and-socket joints ensure good alignment. The lower grip holds a
container that plays the role of a dead load. The load can be increased or decreased at a well-controlled rate by supplying or removing water by
means of a pump. The strain is measured by a strain gauge attached to the sample. Since the stress needed to induce the MT is very sensitive to
temperature, it is important to make sure that equivalent experiments are carried out at the same temperature. For this reason we use a cryofurnace
(which enables temperature control to an accuracy of 0.1 K.) that can be adapted to both devices. In addition, in order to be able to compare the
hysteresis loops obtained with the two devices, the same strain gauge was used in all experiments and the load cell for each set-up was calibrated
using standard weights. For selected experiments at room temperature (without the cryofurnace), an optical microscope was incorporated into both
devices, which enables in situ observations of microstructural changes.
\begin{figure}[htb]
\begin{center}
  \epsfig{file=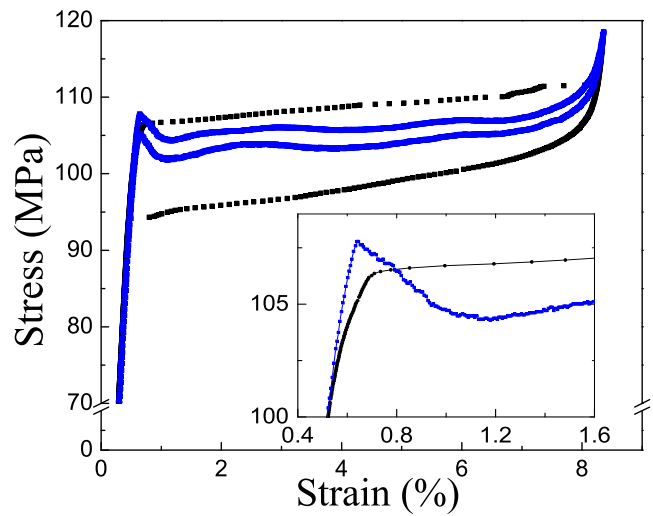,width=8.5cm,clip=}
\end{center}
\caption{\label{FIG2}  (Color online) Experimental stress-strain
hysteresis loop at 303.1 K in a Cu$_{68}$Zn$_{16}$Al$_{16}$ single
crystal, obtained in strain-driven (squares) experiments at 0.005
mms$^{-1}$ and stress-driven (circles) experiments at 0.4
Ns$^{-1}$. The inset shows an enlarged view of the low-strain
region.}
\end{figure}
\begin{figure}[htb]
\begin{center}
  \epsfig{file=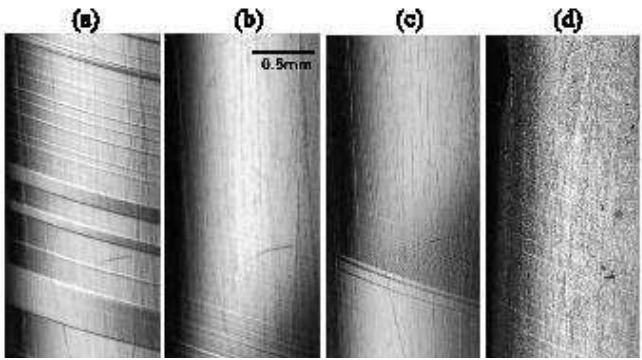,width=8.7cm,clip=}
\end{center}
\caption{\label{FIG3}  Micrographs showing the microstructure
at different stages of the MT: (a) 0.45\% , (b)
0.48 \% and (c) 0.57\% strain. They illustrate the re-entrant behaviour
in  the {\it strain-driven} experiment.
(d) 101.6 MPa, corresponds to the stress-driven case and
illustrates nucleation of thin martensitic plates.}
\end{figure}

\section{Results and Discussion}

Fig.~\ref{FIG2} shows typical results for the stress-strain curves obtained with the two devices. Upon increasing load, the parent cubic phase
persists until it becomes unstable at a given value of the load, and the MT starts. Different hysteresis loops are observed in the two cases. Almost
the entire {\it strain-driven} loop is enclosed within the {\it stress-driven} loop, which has a much larger area. In the {\it stress-driven} case,
the stress slightly increases during the MT with an average slope that decreases as the driving rate is reduced. Extrapolation to zero rate still
gives a finite (small) slope which can be caused by a weak concentration gradient along the axis of the sample. Different behaviour is observed in
the {\it strain-driven} experiment. In this case, once the transition starts, the stress relaxes to a lower value (yield point effect) and then
weakly oscillates around an almost constant value as the MT progresses. For both driving modes, when the transition ends, a new reversible elastic
regime in the martensitic phase is reached. Upon unloading, the behaviour parallels that observed in the loading branch, but with hysteresis. In
particular for the {\it strain-driven} experiment, before the end of the reverse transition, there is an increase in the stress which reproduces the
yield stress effect.

Optical observations have revealed significant differences in the early stages of the MT as shown in Fig.~\ref{FIG3}. The inset in Fig.~\ref{FIG2}
presents an enlarged view of the upper branch of the hysteresis loops on the low-strain region. The yield point in the {\it strain-driven}
experiment is a consequence of re-entrant behaviour within the parent phase, associated with shrinkage and eventually the disappearance of
previously formed martensitic plates. Such re-entrant behaviour is also observed on unloading, which causes the yield point of the lower branch of
the hysteresis loop. The evolution of the microstructure associated with this re-entrant behaviour is illustrated by the micrographs in Fig.~
\ref{FIG3} (a),(b) and (c). The existence of a yield point in conventional {\it strain-driven} experiments has also been reported for other
martensitic alloys \cite{Landa2004,Iadicola2004}. Moreover, computer simulations of the mechanical response of a {\it strain-driven} martensitic
system also show a yield point \cite{Ahluwalia2006}.

For the {\it stress-driven} experiment there is a deviation of the pure elastic behaviour well below the yield stress. This effect is due to the
nucleation of a set of very thin martensitic plates all over the sample, as illustrated by the micrograph in Fig. \ref{FIG3}(d). Interestingly, this
is the only region where the {\it strain-driven} loop is outside the {\it stress-driven} one. Actually, when driving the load, nucleation can occur
at lower values since the system can more freely adapt to shape changes. For both driving mechanisms, steady state growth following the early stages
is macroscopically similar, with parallel interfaces growing towards both ends of the specimen.

It is worth mentioning that the behaviour found in the present experiments for MT seems to be rather common in other systems. For instance,
magnetization-driven magnetic hysteresis loops also show re-entrant behaviour \cite{Grosse1977} while no re-entrancy is present in magnetic
field-driven loops. Furthermore, recent theoretical approaches and computer simulation studies for athermal plastic deformation
\cite{Bouchbinder2006} provide curves with yield points when the control variable is the strain. Interestingly, wiggly trajectories and re-entrance
have even been reported to occur in displacement driven nanoscale systems such as deformation of gold nanowires \cite{Rubio2001} and RNA unfolding
\cite{Manosas2005}.

\begin{figure}[htb]
\begin{center}
  \epsfig{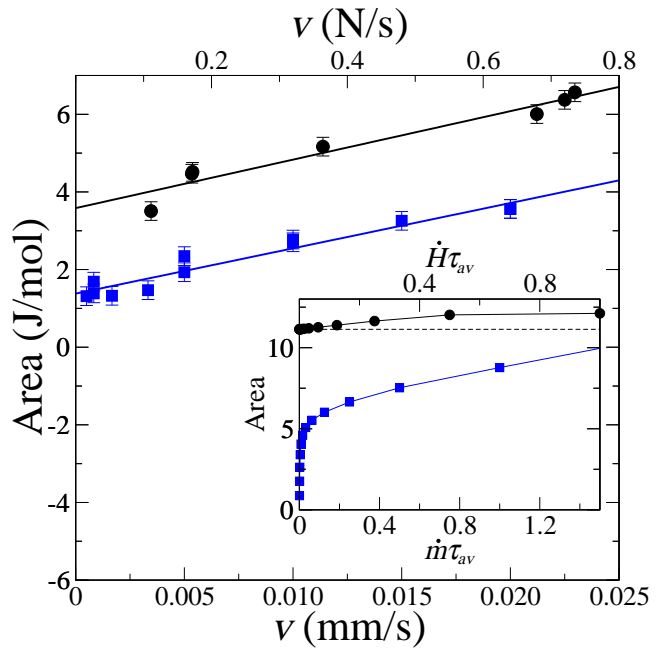}
\end{center}
\caption{\label{FIG4} (Color online) Area of the hysteresis loops as a function
  of the driving rate for the {\it stress-driven} case (circles) and
  the  {\it strain-driven} case  (squares). The inset shows the corresponding results
  obtained by simulation of the T=0 RFIM}
\end{figure}

A relevant quantity in hysteretic systems is dissipated energy which is given by the area enclosed within the hysteresis loops. A noteworthy feature
(see Fig. \ref{FIG2}) is that the dissipated energy is much larger for the {\it stress-driven} case than for the {\it strain-driven} one. The
dissipated energy is expected to depend on the driving rate of the control parameter. We have performed a series of experiments at different rates
and the results of the area of the loops are plotted in Fig. \ref{FIG4} as a function of the rate. The dissipated energy increases with increasing
rate in both cases.  Extrapolation to zero rate with least-squares fits (lines) indicates non-vanishing hysteresis. Similar behaviour has also been
reported for magnetic systems \cite{Nistor05}. It is also important to note that even in the zero-rate limit, the dissipated energy for the {\it
stress-driven} experiment is always larger than for the {\it strain-driven} case.

We now discuss the experimental results presented here in relation to two recent approaches (in magnetic terminology) aimed at analysing $H$-driven
and $M$-driven trajectories in the 3d-RFIM.  Both models predict the existence of a yield point in the $M$-driven loop in agreement with
experiments. In the $H$-driven case the force cannot relax and therefore the system can only reach the closer metastable states by a discontinuity
in the generalitzed displacement. In contrast, for the $M$-driven case the generalized displacement is constrained but the force is able to relax to
lower values in such a way that the system is able to follow a trajectory that approaches equilibrium, thus resulting in a lower dissipation. The
mechanism to relax the force is expected to depend on the details of the specific dynamic features of each system. The present experiments show that
in MT the relaxation is achieved by retransformation. This is also the mechanism in the model in Ref. \cite{Illa2006b}. The wiggly trajectories (see
Fig. \ref{FIG2}) can be interpreted within the framework of the models as being due to the specific distribution of disorder in the system.  One of
the models \cite{Illa2006a} uses $T=0$ adiabatic (infinitely slow) dynamics for the $M$-driven case which is the analogue of standard adiabatic
dynamics introduced by Sethna et al.  \cite{Sethna1993} in the $H$-driven case. The definition of the measured (output) field assumes that the
external force instantaneously equals the highest value of the internal forces. Consequently, the $M$-driven loops have vanishing area. The
strain-driven experiments presented here indicate that although the area is quite small compared with the force-driven situation, after
extrapolation to zero driving rate, a certain amount of dissipated energy still remains.

We have extended the $T=0$ RFIM to incorporate the effect of a finite driving rate. For the $H$-driven case such an extension was proposed
\cite{PerezReche2004} by assuming an intrinsic average time for the avalanches $\tau_{av}$ to relax. In this case the finite-rate $H$-driven loops
can be constructed from the adiabatic $H$-driven loops by considering the discontinuous step-like behaviour of the driving field with a certain
$\Delta H$ in such a way that ${\dot H} = \Delta H / \tau_{av}$. This method allows the area of the loops as a function of the driving rate to be
computed. Results obtained for a cubic system with size $L=20$ and standard deviation of the random-fields $\sigma=1.0$ are shown in the inset of
Fig.  \ref{FIG4} (circles).  Data correspond to averages over more than 1000 realizations of the random fields. We have considered a similar
definition in the simulations of finite-rate $M$-driven systems. It is assumed that the force is only able to relax at given values of the
magnetization which are separated by $\Delta m$ so that ${\dot m} = \Delta m / \tau_{av}$. An example of the obtained results for the same system is
also shown in the inset of Fig.~\ref{FIG4} (squares). Results from numerical simulations satisfactorily reproduce the major features found
experimentally, i.e. an increase in dissipation with increasing rate, and the fact that the dissipated energy is always larger for the force-driven
cases. Only at low-rate limit do numerical data show the already mentioned tendency towards a vanishing area. Such a tendency is due to the
assumption of the infinitely fast response of the system assumed in the models and therefore it is not corroborated by experimental data. The
finite-rate models exactly predict that dissipation in the large-rate limit of the $M$-driven dynamics reaches the zero-rate (adiabatic) limit of
the $H$-driven case (see inset of Fig.~\ref{FIG4}).
\begin{figure}[htb]
\begin{center}
  \epsfig{file=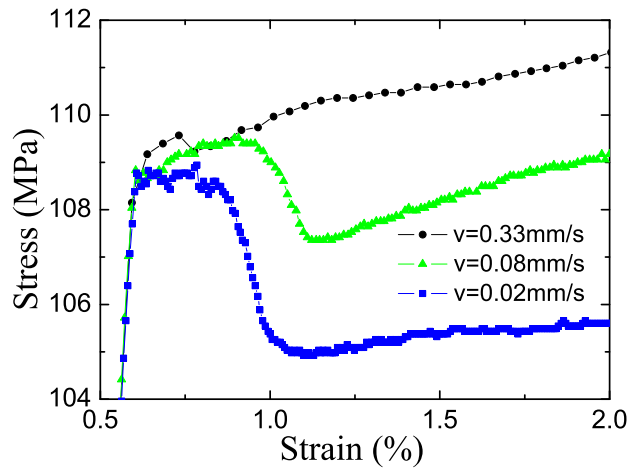,width=8.5cm,clip=}
\end{center}
\caption{\label{FIG5} (Color online) Detail of the rate dependence of the {\sl strain driven}
  loops at high rates, as indicated.}
\end{figure}
Experiments should provide a test of this prediction. As shown in Fig.~\ref{FIG5}, experimental data indicate a tendency to avoid re-entrance as
rate is increased (the yield point is less and less pronounced) in such a way that the corresponding {\it strain-driven} trajectories approach the
zero-rate {\it stress-driven} one. This is in agreement with numerical simulation predictions. However, it must be taken into account that release
(or absorption) of latent heat (which needs to be driven away) associated with the first-order nature of the transition, also contributes to
dissipation as evidenced by the increasing slope with increasing rate in the {\sl strain-driven } curves (see Fig.~\ref{FIG5}). Actually, this extra
effect cannot be reproduced by the model which does not take latent heat into account.

\section{Conclusion}

We have measured the stress-strain curves during the martensitic transition of a Cu-Zn-Al single crystal. Experiments have been conducted under well
controlled force and displacement conditions by means of specifically adapted experimental set-ups. This enables a comparison of the hysteresis
loops obtained under different control variables. It has been shown that metastable trajectories strongly depend on the driving mechanism.
Interestingly, the displacement driven loops exhibit yield point and lower dissipation. In situ optical microscopy has shown that the mechanism for
re-entrance is the re-transformation of part of the material. Results essentially conform to recent theoretical predictions for athermal systems. In
spite of the fact that the microscopic mechanism determining the features of the hysteresis loops can be specific to each particular system, it is
expected that the main results reported in the present study will be common to many ferroic hysteretic systems undergoing athermal first-order phase
transitions.

\acknowledgments This work has received financial support from CICyT (Spain), project MAT2004-1291, CIRIT (Catalonia), project 2005SGR00969, and
Marie-Curie RTN MULTIMAT (EU), contract MRTN-CT-2004-5052226. R.Romero acknowledges a grant from Secretaria de Estado de Universidades e
Investigaci\'on (Spain). The authors acknowledge F.J.P\'erez-Reche, M. Ahlers, T. Lookman, A. Saxena and M.L.Rosinberg for fruitful discussions, and
O.Toscano for experimental assistance.

\vspace{-2mm}


\bibliography{apssamp}

\begin{thebibliography}{99}
\bibitem{feedback} There are tensile machines which enable the load
to be controlled instead of the displacement by making use of a
feedback procedure. However, since the MT
occurs very rapidly (typically at the speed of sound) the feedback
does not ensure a constant load rate at all times.
\end{thebibliography}

\end{document}